\documentclass[11pt]{article}
\usepackage{amsmath,amssymb,color,graphics,epsfig}

\usepackage{amsfonts}

\newcommand{\be}{\begin{equation}}
\newcommand{\ee}{\end{equation}}
\newcommand{\bea}{\setlength\arraycolsep{2pt} \begin{eqnarray}}
\newcommand{\eea}{\end{eqnarray}}

\def\ft#1#2{{\textstyle{\frac{\scriptstyle #1}{\scriptstyle #2} } }}
\def\fft#1#2{{\frac{#1}{#2}}}

\def\0{{\sst{(0)}}}
\def\1{{\sst{(1)}}}
\def\2{{\sst{(2)}}}
\def\3{{\sst{(3)}}}
\def\4{{\sst{(4)}}}
\def\5{{\sst{(5)}}}
\def\6{{\sst{(6)}}}
\def\7{{\sst{(7)}}}
\def\8{{\sst{(8)}}}
\def\sst#1{{\scriptscriptstyle #1}}

\begin{document}

\begin{flushright}
\end{flushright}

\vspace{25pt}
\begin{center}
{\large {\bf Charged Black Holes in Colored Lifshitz Spacetimes \\ }}

\vspace{10pt}
Zhong-Ying Fan and H. L\"u

\vspace{10pt}

{\it Department of Physics, Beijing Normal University, Beijing 100875, China}

\vspace{40pt}

\underline{ABSTRACT}
\end{center}

We consider Einstein gravities coupled to a cosmological constant and $SU(2)$ Yang-Mills fields in four and five dimensions.  We find that the theories admit colored Lifshitz solutions with dynamic exponents $z>1$.  We study the wave equations of the $SU(2)$ scalar triplet in the bulk, and find that the vacuum color modifies the scaling dimensions of the dual operators.  We also introduce a Maxwell field and construct exact solutions of electrically-charged black holes that asymptote to the $D=4$, $z=3$ and $D=5$, $z=4$ colored Lifshitz spacetimes.  We derive the thermodynamical first law for general colored and charged Lifshitz black holes.

\vfill {\footnotesize Emails: zhyingfan@gmail.com \ \ \ mrhonglu@gmail.com}

\thispagestyle{empty}

\pagebreak



\newpage

\section{Introduction}

Yang-Mills fields are responsible for two of the four known fundamental interactions in nature.  Their back reactions to the spacetime geometry are much less understood than that of a Maxwell field. Particle-like static solutions carrying Yang-Mills charges were studied in \cite{Bartnik:1988am}.  Through numerical analysis, it was demonstrated \cite{Bizon:1990sr} that colored black holes supported by $SU(2)$ Yang-Mills fields do exist, although they do not carry any global Yang-Mills charges.  The situation becomes better when the theory involves a cosmological constant. Asymptotic anti-de Sitter (AdS) black holes with global Yang-Mills charges have been shown to exist by numerical analysis \cite{Torii:1995wv,Winstanley:1998sn}. Few examples of exact black holes are known except for those supersymmetric ones constructed in supergravities [5-9]. 
Recently, some exact $SU(2)$-colored and charged black holes were obtained in conformal gravity \cite{Fan:2014ixa}, where the thermodynamical first law of AdS black holes carrying the global Yang-Mills charges were also derived.

   In this paper, we consider four and five-dimensional Einstein gravities with a cosmological constant minimally coupled to the $SU(2)$ Yang-Mills fields.  We find that the theories admit colored Lifshitz vacua with the scaling exponents $z>1$.  The Lifshitz solutions are the generalizations of the AdS spacetime, with the metric \cite{Kachru:2008yh}
\be
ds^2 = \ell^2 \Big(-r^{2z} dt^2 + \fft{dr^2}{r^2} + r^2 dx_i dx_i\Big)\,.
\ee
It is homogeneous but not Einstein, and is invariant under the scaling $r\rightarrow r/\lambda$, provided that
\be
t\rightarrow \lambda^z t\,,\qquad x_i \rightarrow \lambda x_i\,.
\ee
It reduces to AdS when $z=1$. These backgrounds were proposed \cite{Kachru:2008yh} as gravity duals of condensed matter systems that exhibit critical behavior of different dynamic scalings between temporal and spatial directions in fixed points.  Although the theory \cite{Kachru:2008yh} proposed for the Lifshitz backgrounds appeared to be ad hoc, the vacua were subsequently found in a variety of theories, including strings and M-theory, as well as some further ad hoc theories [10-42].  Many examples of Lifshitz black holes have also been found [26-42].

We find it is rather intriguing that Lifshitz spacetimes themselves can be colored in Einstein-Yang-Mills gravity.  Furthermore, we find that introducing a Maxwell field allows us to construct some exact charged black holes in the colored Lifshitz backgrounds.  The paper is organized as follows.  In section 2, we present the theories and derive the equations of motion.  In section 3, we construct $SU(2)$-colored Lifshitz spacetimes and study the Green's function of operators dual to an $SU(2)$ scalar triplet in the bulk.  In section 4, we construct exact charged black holes in the $D=4$, $z=3$ and $D=5$, $z=4$ Lifshitz spacetimes.  In section 5, we derive the thermodynamical first law of our general black holes and show that the Yang-Mills hair do not contribute to the first law.  In section 6, we embed the $D=4$ theory in gauged supergravity that can be lifted to $D=11$.  We conclude the paper in section 7.

\section{The theory and equations of motion}

We start with $D$-dimensional Einstein gravity with a cosmological constant minimally coupled to $SU(2)$ Yang-Mills fields $A_\mu^a$ ($a=1,2,3$) and Maxwell field ${\cal A}_\mu$.  The Lagrangian is
\be
{\cal L}_D=\sqrt{-g} \Big(R-2\Lambda-\frac{1}{2g_s^2}F^a_{\mu\nu}F^{a\mu\nu} - \fft{1}{2} {\cal F}_{\mu\nu} {\cal F}^{\mu\nu}\Big)\,.
\ee
The Yang-Mills and Maxwell field strengths are defined as
\be
F^a_{\mu\nu}=\partial_\mu A^a_\nu-\partial_\nu A^a_\mu+\epsilon^{abc}A^b_\mu A^c_\nu\,,\qquad
{\cal F}_{\mu\nu}=\partial_\mu {\cal A}_\nu - \partial_\nu {\cal A}_\mu\,.
\ee
The full set of covariant equations of motion are:
\bea
&&R_{\mu\nu} = \ft{2\Lambda}{D-2}\,g_{\mu\nu} + \ft{1}{g_s^2}(g^{\rho\sigma}F^a_{\mu\rho}F^a_{\nu\sigma}-\ft1{2(D-2)} F^2 g_{\mu\nu}) + g^{\rho\sigma}{\cal F}_{\mu\rho}{\cal F}_{\nu\sigma}-\ft 1{2(D-2)} {\cal F}^2 g_{\mu\nu}\,,\cr
&&\nabla_\mu F^{a\mu\nu}+\epsilon^{abc}A_\mu^b F^{c\mu\nu}=0\,,
\qquad \nabla_\mu {\cal F}^{\mu\nu}=0\,.\label{abseom}
\eea
In this paper, we focus on $D=4, 5$ only and consider a class of solutions whose metrics take the form
\be
ds^2 = -h(r) dt^2 + \fft{dr^2}{f(r)} + r^2 dx_i dx_i\,,
\ee
where $i=1,2$ for $D=4$ and $i=1,2,3$ for $D=5$.  The Ansatz for the Maxwell field is electric, i.e.~${\cal A}=\phi(r) dt$. The Maxwell equation implies that $
\phi' = \fft{q}{r^{D-2}}\, \sqrt{h/f}$. In this paper, a prime denotes a derivative with respect to $r$.
The Ansatz for the $SU(2)$ Yang-Mills fields in $D=5$ is given by
\be
A \equiv \sum_{a=1}^3 \tau^a A^a =\psi(r) (\tau^1 dx_1 + \tau^2 dx_2 + \tau^3 dx_3)\,,
\ee
where $\tau^a$ are the Pauli matrices.  The Ansatz for $D=4$ is given by simply dropping the $dx_3$ term in the above, which implies $A^3=0$.

Substituting the full Ans\"atze into (\ref{abseom}), we find that the full set of equations of motion are now reduced to three independent differential equations
\bea
&&\psi'' + \Big(\fft{h'}{2h} + \fft{f'}{2f} + \fft{D-4}{r}\Big)\psi' - \fft{\psi^3}{r^2 f}=0\,,\qquad \fft{f'}{f} - \fft{h'}{h} + \fft{2\psi'^2}{g_s^2 r}=0\,,\cr
&& \fft{f'}{h'} + \fft{h'}{h} + \fft{(D-3) \psi^4}{g_s^2 r^3 f} +
\fft{2q^2}{(D-2) r^{2D-5} f} + \fft{4\Lambda r}{(D-2) f} + \fft{2(D-3)}{r}=0\,.
\label{geneom}
\eea
Note that the three equations are valid only for $D=4,5$.

\section{Colored Lifshitz vacua}

The self-interaction of the Yang-Mills fields allows us to find a class of Lifshitz vacua with a generic exponent $z>1$:
\be
h=\ell^2 r^{2z}\,,\qquad f=\fft{r^2}{\ell^2}\,,\qquad
\psi=\sqrt{\fft{z+D-3}{(D-3)\ell^2}}\, r\,,\qquad \phi=0\,,\label{psi1}
\ee
with
\be
\Lambda = -\fft{(D-2)(z^2 + (D-2)z+(D-1))}{4\ell^2}\,,\qquad g_s^2=\fft{z+D-3}{(D-3)(z-1)\ell^2}\,.
\ee
Note again that the solutions are valid for $D=4,5$ only.  The $D=4$ solution was obtained in \cite{Devecioglu:2014iia}.

Since it is rather unusual to have a colored vacuum, we would like to study the effect of its color on the boundary field theory.  After performing the scaling $x_i\rightarrow \ell x_i$, the solutions take the form
\bea
ds^2 &=& \ell^2\Big(-r^{2z}dt^2+\frac{dr^2}{r^2 }+r^2(dx_1^2+dx_2^2 + dx_3^2)\Big)\,,\cr
A &=&p r\,(\tau^1 dx_1+\tau^2 dx_2 + \tau^3 dx_3)\,,\qquad p=\sqrt{\ft{z+D-3}{D-3}}\,.\label{lifsol}
\eea
For $D=4$, we simply drop the $dx_3$ terms everywhere.  (Note that $p$ is independent of $\ell$ owing to the scaling of $x_i$.) It is clear that a minimally-coupled massive scalar is colorblind.  We consider instead a minimally-coupled scalar triplet under the $SU(2)$, namely
\be
{\cal L}_{\phi}=\sqrt{-g}\big(-\ft12 D_\mu \phi^a D^{\mu}\phi^a-\ft12 m^2 \phi^a\phi^a\big)\,,
\ee
where $D_\mu \phi^a=\partial_\mu \phi^a+\epsilon^{abc}A^b_\mu \phi^c$.  The covariant equations are
\be
\frac{1}{\sqrt{-g}}D_\mu(\sqrt{-g}g^{\mu\nu}D_\nu \phi^a)-m^2\phi^a=0\,.
\ee
We find that the scalar equations can be solved using the Ansatz
\be
\phi^a(r,t,x^i)=k_a\,\varphi(r)\,e^{-{\rm i}\omega t+{\rm i} (k_1 x_1+k_2 x_2 + k_3 x_3)}\,,
\ee
where $\varphi$ satisfies
\be
\varphi''+\frac{z+D-1}{r}\varphi'+\Big(\frac{\omega^2}{r^{2z+2}}-
\frac{k^2}{r^4}-\frac{m_{\rm eff}^2\ell^2}{r^2}\Big)\varphi=0\,,
\ee
where $k^2=k_ik_i$ and $m_{\rm eff}^2=m^2+(D-3) p^2/\ell^2$. (There is no $k_3$ for $D=4$.) Interestingly the magnetic Yang-Mills configuration (\ref{lifsol}) contributes a positive constant shift to the mass square of the scalar triplet.

In the asymptotic $r\rightarrow \infty$ limit, the $(\omega, k)$ terms are both subdominant, and we have:
\be
\varphi\rightarrow \frac{A}{r^{\Delta_{-}}}+\frac{B}{r^{\Delta_{+}}},\qquad \Delta_\pm=\ft12(z+D-2)\pm \nu\,,
\ee
where $\nu=\ft12\sqrt{(z+D-2)^2+4m_{{\rm eff}}^2\ell^2}$. In standard quantization, the vev of the dual boundary operator is defined by $\langle O_+ \rangle=B$, which is a normalizable mode of scaling dimension $\Delta_+$. The two-point function can be derived as $G_{O_+}=B/A$. Thus for the operators to have real dimensions, we need to impose a Breitenlohner-Freedman (BF)-type bound, namely
\be
m^2\geq m_{BF}^2\,,\qquad \hbox{with}\qquad
m_{BF}^2= -\frac{(z+D-2)^2+4(D-3) p^2}{4\ell^2}\,.
\ee
The color parameter $p$ of the vacuum makes the BF bound more negative. For the mass range in $0\leq m^2-m_{BF}^2\leq 1/\ell^2$, the mode $A$ is also normalizable, in which case, one may also work in the alternative quantization with $\langle O_- \rangle=A$, and $G_{O_-}=A/B$.

In the following, we shall work in standard quantization. For vanishing $\omega$, the equation (16) can be analytically solved in terms of a Bessel function $\varphi=r^{-\frac{z+D-2}{2}} K_\nu(k/r)$. Taking the boundary limit, we find
\be
\varphi \rightarrow \fft{k^{-\nu}\, \Gamma(\nu)}{2^{1-\nu} r^{\Delta_-}}+
\fft{k^\nu\, \Gamma(-\nu)}{2^{1+\nu} r^{\Delta_+}}\,.
\ee
The Green's function for $\omega=0$ can be immediately read off. Analogously, we can also obtain the Green's function for $k=0$.  The results are
\be
G_{O_+}(\omega=0,k)=(\ft12k)^{2\nu} \fft{\Gamma(-\nu)}{\Gamma(\nu)}\,,\qquad
G_{O_+}(\omega,k=0)=\Big(\fft{{\rm i}\omega}{2z}\Big)^{2\nu} \fft{\Gamma(-\fft{\nu}{z})}{\Gamma(\fft{\nu}{z})}\,.
\ee
For generic $(\omega,k)$, the equation (16) cannot be solved analytically except for the relativistic case $z=1$ for which the Yang-Mills fields decouple from the vacuum.

\section{Charged black holes in colored Lifshitz spacetimes}

Having obtained the $SU(2)$-colored Lifshitz vacua in Einstein-Yang-Mills theories in four and five dimensions, we would like to construct black holes that are asymptotic to these vacua.  We first present an exact solution for $z=3$ in $D=4$, given by
\bea
ds^2_4&=&\ell^2\Big(-r^{6} \tilde f dt^2+\frac{dr^2}{r^2 \tilde f}+r^2(dx_1^2+dx_2^2)\Big)\,,\qquad
\tilde f=1-\frac{q^2\ell^2}{2r^4}\,,\cr
A&=&2r(\tau^1 dx_1+\tau^2 dx_2)\,,\quad {\cal A}=(\phi_0 +\ell^2 q r) dt\,,\quad
g_s^2=\fft{2}{\ell^2}\,,\quad \Lambda=-\fft{9}{\ell^2}\,.
\eea
Note that it is straightforward to add a magnetic charge to ${\cal A}$. We also obtain an exact solution in $D=5$ for $z=4$:
\bea
ds_5^2 &=& \ell^2\Big(-r^{8} \tilde f dt^2+\frac{dr^2}{r^2 \tilde f}+r^2(dx_1^2+dx_2^2 + dx_3^2)\Big)\,,\qquad
\tilde f=1-\frac{q^2\ell^2}{3r^6}\,,\cr
A&=&\sqrt3\, r(\tau^1 dx_1+\tau^2 dx_2 + \tau^3 dx_3),\,\,\,\, {\cal A}=(\phi_0 +\ell^2 q r) dt,\,\,\,\,
g_{s}^2=\fft{1}{\ell^2},\,\,\,\, \Lambda=-\fft{24}{\ell^2}\,.
\eea
In both cases, $\phi_0$ is a constant which can be viewed as the gauge parameter or simply as an integration constant.  We shall choose a gauge such that ${\cal A}$ vanishes on the horizon.  The solutions describe black holes with event horizons at $r=r_0>0$ where $\tilde f(r_0)=0$.  The temperature and entropy are given by
\be
T=\ft{1}{2\pi}(D-2) r_0^{D-1},\qquad S=\ft14\omega(\ell r_0)^{D-2}\,.\label{specialts}
\ee
where $\omega=\int dx_1 dx_2$ for $D=4$ and $\omega=\int dx_1 dx_2 dx_3$ for $D=5$.  The electric charge can be easily obtained; however,
the electric potential $\Phi$, the thermodynamical conjugate to the charge, is rather tricky in this case.  The usual definition of $\Phi$ is the difference of $\phi$ between the horizon, which we set zero, and its value at the asymptotic infinity, which is divergent for our solutions. This phenomenon was seen already in \cite{Liu:2014dva}. We shall follow \cite{Liu:2014dva} and  show in the next section that there is still a way to define the first law of thermodynamics for these black holes.

\section{The thermodynamics of general colored black holes}

\subsection{The structure of a general black hole}

Although we are not able to solve the differential equations (\ref{geneom}) to obtain the general analytical solutions, we can nevertheless study the structure of the black hole solutions and obtain their thermodynamical first law.  For a solution to be a black hole, there must be an event horizon, which we assume to be located at $r=r_0>0$.  The near horizon geometry can be obtained by the Taylor expansions around $r=r_0$:
\bea
h&=& \tilde{a}_1(r-r_0)+\tilde{a}_2(r-r_0)^2+ \cdots\,,\cr
f &=& \tilde{b}_1(r-r_0)+\tilde{b}_2(r-r_0)^2+ \cdots\,,\cr
\psi &=& \tilde{c_0}+\tilde{c}_1(r-r_0)+\tilde{c}_2(r-r_0)^2+ \cdots\,.\label{nearhorizon}
\eea
Substituting these into the equations of motion (\ref{geneom}), we can solve them order-by-order and all the coefficients of the Taylor expansions can be expressed in terms of $(r_0, \tilde c_0, q)$ and $\tilde a_1$.  The parameter $\tilde a_1$ represents the freedom of arbitrarily scaling the time coordinate.
It has to be fixed so that the time at asymptotic infinite satisfies the standard convention.  Thus the general horizon geometry is specified by three non-trivial parameters.

   In order to see whether such a horizon can be integrated out smoothly to infinity, we study how the asymptotic structure deviates from the Lifshitz vacua.  To do this, we let
\be
h = r^{2z} (1 + h_1)\,,\qquad f=r^2 (1+ f_1)\,,\qquad
\psi=p\, r\, (1 + \psi_1)\,.
\ee
Without loss of generality, we have set $\ell=1$ for simplicity. The functions $h_1$,$f_1$ and $\psi_1$ can be solved at the linear level, given by
\bea
h_1&=&\lambda_0 - \fft{m}{r^{z+n}} + \fft{a_-\psi_-}{r^{\fft12(z+n-\sigma)}} +
\fft{a_+\psi_+}{r^{\fft12(z+n+\sigma)}} - \fft{c_1 q^2}{r^{2n}}\,,\cr
f_1&=&\fft{\lambda_1 m}{r^{z+n}} -\fft{2(z-1)\psi_-}{r^{\fft12(z+n-\sigma)}}-
\fft{2(z-1)\psi_+}{r^{\fft12(z+n+\sigma)}} - \fft{c_2 q^2}{r^{2n}}\,,\cr
\psi_1 &=& \fft{\lambda_2 m}{r^{z+n}} +\fft{\psi_-}{r^{\fft12(z+n-\sigma)}}+
\fft{\psi_+}{r^{\fft12(z+n+\sigma)}} - \fft{c_3 q^2}{r^{2n}}\,,\label{linear}
\eea
where $n=D-2$. For $D=4$, we have $\sigma^2=9z^2+4z+12$, and
\bea
&&a_\pm = \ft{2(z-1)(z-2\pm\sigma)}{z+2\pm\sigma}\,,\qquad
\lambda_1 = -\ft{z(z+2)}{z^2 + 2}\,,\qquad
\lambda_2 = \ft{z+2}{2(z+1)(z^2+2)}\,,\cr
&&\{c_1,c_2,c_3\} = \ft{1}{(z-1)(z-2)(z+3)} \{\ft34 (z^2-1), 3 (z-1),
\ft14 (z-3)\}\,;
\eea
for $D=5$, we have $\sigma=9z^2 + 14 z + 17$, and
\bea
&&a_\pm =\ft{2(z-1)(z-1\pm\sigma)}{z+3\pm \sigma}\,,\quad
\lambda_1=-\ft{z(z+3)}{z^2+z+2}\,,\quad
\lambda_2=\ft{z+3}{2(z+2)(z^2+z+2)}\,,\cr
&&\{c_1,c_2,c_3\}=\ft{1}{(z-3)(z^2+4z-8)}\{ \ft19 (z+2)(5z-8), 3z-4, \ft16(z-4)\}\,.
\eea
Thus we see at the linear level, the solutions contain parameters $(\lambda_0,m)$ associated with graviton modes, $(\psi_-,\psi_+)$ associated with Yang-Mills hair and $q$ associated with electric charge. The $\lambda_0$ represents the freedom of arbitrarily scaling the time coordinate and hence we set $\lambda_0=0$ to keep our convention on time. Since $z>1$, the $\psi_-$ terms are all divergent asymptotically, and hence must be excluded.
This is analogous to colored black holes in $D=4$ Minkowski spacetime and the exclusion of $\psi_-$ terms implies the absence of the global Yang-Mills charges \cite{Bizon:1990sr}. We expect that the fine tuning of the horizon parameters to get rid of the $\psi_-$ terms has the effect that the black hole solutions have two non-trivial parameters, the mass and the electric charge.  Indeed, the numerical analysis for $q=0$ and $D=4$ was performed in \cite{Devecioglu:2014iia}.

\subsection{Thermodynamical first law}

Having analysed the structure of the general colored black hole solutions, we now adopt the Wald formalism \cite{wald1,wald2} to derive their thermodynamical first law. The procedure has been performed for a variety of Lifshitz black holes in \cite{Liu:2014dva}.  The advantage of this procedure is that one can obtain the first law without knowing an exact solution; the behaviors on the horizon and at the asymptotic infinity suffice.  The Wald formula, specialized on to static black holes for general Proca fields and $SU(2)$ Yang-Mills fields, was derived in \cite{Liu:2014tra,Fan:2014ixa} respectively.  Combining the two results together, we have
\be
\delta H= -\fft{\omega}{16\pi} r^n\, \sqrt{\fft{h}{f}}\, \Big( \fft{n}{r}
\delta f + \fft{4}{g_s^2 r^2} f \psi' \delta \psi +
\fft{2f}{h} \phi \delta \phi' +  \phi \phi' \Big(\fft{\delta f}{h} - \fft{f\delta h}{h^2}\Big)\Big)\,,\label{deltaH}
\ee
where $n=D-2$ and $\delta$ acts on the integration constants of a solution. The first law is a consequence of $\delta H_+=\delta H_\infty$.  Substituting the near-horizon solution (\ref{nearhorizon}) into (\ref{deltaH}), and then taking the limit $r\rightarrow r_0$, we find
\be
\delta H_+ = T \delta S\,,\quad\hbox{with}\quad
T=\fft{\sqrt{h'(r_0) f'(r_0)}}{4\pi}\,,\qquad
S=\fft{1}{4}\omega r_0^n\,.
\ee
Note that a gauge choice $\phi(r_0)=0$ was made for the above result.  Substituting the linear solution (\ref{linear}) with $\psi_-=0$ into (\ref{deltaH}) and taking the $r\rightarrow \infty$ limit, we find
\be
\delta H_\infty = \delta M - \fft{\omega}{8\pi} \phi_0 \delta q\,,
\quad \hbox{with}\quad
M=
\left\{
  \begin{array}{ll}
    \fft{(z+2)(z^2+1)\omega }{8\pi (z+1)(z^2+2)}\, m\,, \quad& \hbox{for}\qquad D=4, \\
    \fft{(z+3)(3z^2+4z+2)\omega }{16\pi (z+2)(z^2+z+2)}\, m\,, \quad& \hbox{for}\qquad D=5.
  \end{array}
\right.
\ee
It follows that the first law of thermodynamics is given by
\be
dM = TdS + \Phi dQ\,,\label{fl}
\ee
where
\be
\Phi=\phi_0\,,\qquad Q=\fft{1}{8\pi}\int_{r\rightarrow \infty} {*F}=\fft{\omega}{8\pi}q\,.
\ee
Although the first law was derived from a linear analysis, we expect that it is correct for the full non-linear solution. This is because that the first law is bilinear, and hence the coefficients in the first law should be determined by the linear analysis already.  The absence of the contribution of Yang-Mills hair to the first law is understandable since although $\psi_+$ can survive in a solution, its thermodynamical conjugate $\psi_-$ has to be eliminated.  By contrast, colored AdS black holes may have non-vanishing global Yang-Mills charges and the first law contain a term $P^{\rm YM} dQ^{\rm YM}$ as well \cite{Fan:2014ixa}.

As observed previously, the Wald formula simply picks up the $\phi_0$, an integration constant or pure gauge from the solution.  It acquires the interpretation of the potential difference between the horizon and the asymptotic infinity if and only if $\phi(r)$ converges.  To conclude,  the general charged and colored Lifshitz black holes contain two independent parameters, the mass $M$ and the electric charge $Q$.  The Yang-Mills hair represented by $\psi_+$ must be some appropriate function of $M$ and $Q$ fixed by the fine tuning on the horizon.

We can now derive the first law of both exact black holes constructed in section 4.  It contains only one non-trivial parameter $Q$.  The first law is
\be
TdS + \Phi dQ=0\,.
\ee
where $\Phi=\phi_0=-qr_0$.  Interestingly, charged massless Lifshitz black holes with non-vanishing temperature and entropy are quite ubiquitous \cite{Zingg:2011cw,Liu:2014dva,Fan:2014ixa,Fan:2014ala}.

\section{Embedding in gauged supergravity}

In four dimensions, the Einstein-Yang-Mills theory with a cosmological constant can be supersymmetrized to become ${\cal N}=2$ gauged supergravity.  The theory can be lifted to eleven-dimensional supergravity \cite{Pope:1985bu}.  The supersymmetry requires that $\Lambda$ and $g_s^2$ be related by $\Lambda=-6g_s^2$, implying that the Lifshitz solution has an irrational $z=1 + \sqrt6$.

\section{Conclusions}

In this paper, we constructed colored Lifshitz spacetimes with dynamic exponents $z>1$ in four and five-dimensional Einstein gravities coupled to a cosmological constant and $SU(2)$ Yang-Mills fields.  We studied the wave equations of an $SU(2)$ scalar triplet and derived the explicit Green's function of the dual operators for vanishing frequency $\omega$ or momentum $k$.  We found that the spacetime color increased the scaling dimensions of the dual operators by constant shifting the mass square of the scalar triplet in the bulk.

We also introduced a Maxwell field and constructed exact charged black holes for $D=4$, $z=3$ and $D=5$, $z=4$.  We analysed the structure of the general Lifshitz black holes, and showed that they had two non-trivial parameters, the mass $M$ and electric charge $Q$, but with no global Yang-Mills charges.  The Yang-Mills hair parameter $\psi_+$ is a function of $M$ and $Q$. We derived the first law (\ref{fl}) of thermodynamics of the black holes and showed that the Yang-Mills hair did not modify it.

The colored Lifshitz vacua and black holes we constructed provide new tantalizing gravitational backgrounds for studying some condensed matter physics.

\section*{Acknowledgement}

Z.-Y.~Fan is supported in part by NSFC Grants NO.10975016, NO.11235003 and NCET-12-0054; The work of H.L.~is supported in part by NSFC grants NO.11175269, NO.11475024 and NO.11235003.

\end{document}